\documentclass[prd,aps,floats,preprint,preprintnumbers]{revtex4}
\usepackage{amsmath,amssymb}
\usepackage{graphics}
\usepackage{graphicx}
\usepackage{epsfig}

\newcommand {\eqv}{\equiv}
\newcommand {\td}{\tilde}

\newcommand {\ra}{\rightarrow}
\newcommand {\Ra}{\Rightarrow}

\newcommand {\del}{\partial}
\newcommand {\Ld}{\Lambda}
\newcommand {\ld}{\lambda}

\newcommand {\al}{\alpha}

\newcommand {\bfr}{\begin{flushright}}
\newcommand {\efr}{\end{flushright}}
\newcommand {\bfl}{\begin{flushleft}}
\newcommand {\efl}{\end{flushleft}}
\newcommand {\nn} {\nonumber}

\newcommand {\txt}{\textrm}
\newcommand {\bd}{
\begin{document}}
\newcommand {\ed}{\end{document}}

\newcommand {\be}{\begin{equation}}
\newcommand {\ee}{\end{equation}}
\newcommand {\bea}{\begin{eqnarray}}
\newcommand {\eea}{\end{eqnarray}}
\newcommand {\ba}{\begin{array}}
\newcommand {\ea}{\end{array}}

\newcommand{\bbib}{\begin{thebibliography}{99}}
\newcommand{\ebib}{\end{thebibliography}}
\newcommand {\bab}{\begin{abstract}}
\newcommand {\eab}{\end{abstract}}

\newcommand {\bc}{\begin{center}}
\newcommand {\ec}{\end{center}}
\newcommand {\bit}{\begin{itemize}}
\newcommand {\eit}{\end{itemize}}
\newcommand {\ul}{\underline}
%\newcommand{\bcmt}{\begin{comment}}
%\newcommand{\ecmt}{\end{comment}}
%%%%%%%%%%%%%%%%%%%%%%%%%%%%%CCCCCCCCCCCCCCCCCCCCCcolor
\newcommand {\txtc}{\textcolor}

%%%%%%% END NEW COMMANDS  %%%%%%% "functions"
\newcommand {\ad}{\textrm{ad}}
\newcommand {\Ad}{\textrm{Ad}}
% {\cal }
\def\A{{\cal A}}\def\B{{\cal B}}\def\C{{\cal C}}\def\D{{\cal D}}\def\E{{\cal E}}\def\F{{\cal F}}\def\G{{\cal G}}\def\H{{\cal H}}\def\I{{\cal I}}
\def\J{{\cal J}}\def\K{{\cal K}}\def\L{{\cal L}}\def\M{{\cal M}}\def\N{{\cal N}}\def\O{{\cal O}}\def\P{{\cal P}}\def\Q{{\cal Q}}\def\R{{\cal R}}
\def\S{{\cal S}}\def\T{{\cal T}}\def\U{{\cal U}}\def\V{{\cal V}}\def\W{{\cal W}}\def\X{{\cal X}}\def\Y{{\cal Y}}\def\Z{{\cal Z}}

\def\xbar{\bar{x}}\def\ybar{\bar{y}}\def\zbar{\bar{z}}\def\kbar{\bar{k}}\def\pbar{\bar{p}}
%%%%%%%%%%%%%%%%%%%%%%%%%%%%%%%%%%%%%%%%%%%%%%%%%%%%%%%%%%%%%%%%%%%%%%%%%%%%%%%%%%%%%%%%%%%%%%%%%%%%%%%%%%%%%%%%%%%%%%%%%%%%%
% { \color{red} 'hfh'},  \textcolor{red}{gfh}
% {\bf }, {\it }, \underline{}, {\huge{} }, {\large {} },
% \chapter{}, \section{}, \subsection{}, \titlepage{}, \pagebreak{},
%\lable{'label'}, \ref{'label'}, ...
%%%%%%%%%%%%%%%%%%%%%%%%%%%%%%%%%%%%%%%%%%%%%%%%%%%%%%%%%%%%%%%%%%%%%%%%%%%%%%%%%%%%%%%%%%%%%%%%%%%%%%%%%%%%%%%%%%%%%%%%%%%%%
 %\begin{thebibliography}{99}
   %\bibitem{'label'} 'item'
   %\bibitem{'label'}  'item'
 %\end{thebibliography}
 %\cite{'label'}
%%%%%%%%%%%%%%%%%%%%%%%%%%%%%%%%%%%%%%%%%%%%%%%%%%%%%%%%%%%%%%%%%%%%%%%%%%%%%%%%%%%%%%%%%%%%%%%%%%%%%%%%%%%%%%%%%%%%%%%%%%%%%
 %\begin{itemize}
  %\item[1)] 'item'
  %\item[2)] 'item'
 %\end{itemize}
%%%%%%%%%%%%%%%%%%%%%%%%%%%%%%%%%%%%%%%%%%%%%%%%%%%%%%%%%%%%%%%%%%%%%%%%%%%%%%%%%%%%%%%%%%%%%%%%%%%%%%%%%%%%%%%%%%%%%%%%%%%%%
 %\bea x=\left\{\ba{ll}  a& aa\\ c&cc\\d&dd  \ea\right. \eea
 %\bea x=\left\{\ba{ll}  a& aa\\ c&cc\\d&dd  \ea\right\} \eea
 %\bea x=\left(\ba{ll}  a& aa\\ c&cc\\d&dd  \ea\right) \eea
%%%%%%%%%%%%%%%%%%%%%%%%%%%%%%%%%%%%%%%%%%%%%%%%%%%%%%%%%%%%%%%%%%%%%%%%%%%%%%%%%%%%%%%%%%%%%%%%%%%%%%%%%%%%%%%%%%%%%%%%%%%
\bd
\preprint{SU-4252-896 \vspace{2cm}}
\title{Generation and propagation of a q-deformed type of $d^N\neq0$ curvature}
\vspace{2cm}
\author{E. Akofor}\thanks{eakofor@phy.syr.edu}
\vspace{2cm}
\affiliation{Department of Physics, Syracuse University,
Syracuse, NY 13244-1130, USA\vspace{2cm}}
\begin{abstract}
\vspace{1cm}
We present an expression for curvature with q-deformed calculus such as considered in \cite{d-k,b-b-k,f-m-r-s-w}. By exploiting the persistence of Bianchi's second identity, we suggest a way to attach physical meaning to the $q$ parameters and $d^N\neq 0$ condition by introducing a physical current, an example of which may be obtained by a procedure outlined in \cite{akofor}.
\end{abstract}
\maketitle
\section{Introduction}
In this section we explain notation used in the following sections. The appearance of the ``constant'' deformation parameter $q$ is only formal since its index structure in a coordinate basis $\{dx^\mu\}$ of differential $1$-forms can be involved, where general indices will be denoted by $i,j,k,l,...$ meanwhile spacetime indices will be denoted by $\mu,\nu,\al,\beta,...$. In order to illustrate this point, we will show the explicit form of the q-symmetrization in the case of the 2-form of the electromagnetic field in section \ref{q-symm}. $\M$ will be an n-dimensional differentiable manifold and $T(\M)$ will be the tangent bundle over $\M$. The differential forms will be regarded as belonging to the algebra $\Ld_q$ of sections of the $q$-symmetrized tensor algebra $\T^\ast(\M)$ of the linear functionals ~$T^\ast(\M)=\{u:T(\M)\ra \mathbb{C}\}$~ on $T(\M)$. That is,
\bea
&&\T^\ast(\M)= \mathbb{C}\oplus T^\ast(\M)\oplus T^\ast(\M)^{2\otimes_q}\oplus...\oplus T^\ast(\M)^{n\otimes_q},\nn\\
&&T^\ast(\M)^{k\otimes_q}=T^\ast(\M)\otimes_q T^\ast(\M)^{(k-1)\otimes_q},\nn\\
&&\Ld_q=\{f:\M\ra \T^\ast(\M)\}=\Ld_q^0\oplus\Ld^1_q\oplus\Ld^2_q\oplus...\oplus\Ld^n_q,\nn\\
&&\Ld^k_q=\{f:\M\ra T^\ast(\M)^{k\otimes_q}\},\nn\\
&&\Ld_q^0=\F(\M)=\{f:\M\ra \mathbb{C}\}.
\eea
Differentiation is from the left and whenever the q-deformed exterior differential $d$ passes over a q-deformed $j$-form from the left, a factor of $q^j$ is piked up. That is if $f\in \Ld^{|f|}_q,~g\in \Ld^{|g|}_q$  then
\bea
&&d:\Ld_q\ra \Ld_q,~~\Ld^k_q\ra \Ld_q^{(k+|d|)~mod~n},~~d(fg)=df~g+q^{|f||d|}f~dg,
\eea
where we take $|d|=1$. $\omega$ will denote a 2-array (or matrix) of q-deformed differential 1-forms.

We also adopt the following index summation convention which is most important for section \ref{q-symm}. An index is summed over only if it appears more than once (twice, thrice etc) on one side of an equation but does not appear on the opposite side of the equation. For example, the indices $i$ and $j$ in $a_{ij}=u_iv_{ij}$ are not summed over. Similarly, $i,j$ are not summed over but $k,l$ are summed over in $a_{ij}=q_{kj}b_{kl}c_{lik}d_j$. There are exception which should be clear from context. Examples of such exceptions include the assignment of values to the components of an array, eg. in $a_{ij}=2~~\forall i,j$, and in the notation for a matrix $M=(M_{ij})=(\ld_i\al_{ij})$ or a vector $v=(v_i)=(\ld_i u_i)$ or any other array.

\section{Connection and curvature}

Consider a tangent vector field $v=\td{v}^i\del_i:\mathcal{M}\ra T(\mathcal{M})$ in the coordinate basis $\{\del_i\}$ with $d\del_i=\omega_i{}^j\del_j$~(or simply $d\del=\omega\del$)~ and ~$\omega=(\td{\omega}_i{}^j{}_\al dx^\al)$ being a 1-form valued matrix or a connection $1$-form. Then one finds that
\bea
&&d^kv=d^k(\td{v}\del)=\sum^k_{r=0}C_q^{(k,r)}d^{k-r}\td{v}~\Omega_r~\del\eqv \sum^k_{r=0}C_q^{(k,r)}d^{k-r}\td{v}~d^r\del~~\in ~\Ld^k_q\otimes T(\M),\nn\\
&&C_q^{(k,r)}={[k]_q!\over [r]_q![k-r]_q!},\nn\\
&&[n]_q!=[n]_q..[3]_q[2]_q[1]_q,\nn\\
&&[n]_q={1-q^n\over 1-q}=1+q+q^2+..+q^{n-2}+q^{n-1},
\eea
where the sequence of curvature forms $(\Omega_r)$ is given recursively by
\bea
&&\Omega_0=1,\nn\\
&&\Omega_1=\omega,\nn\\
&&\Omega_2=d\omega+q\omega^2,\nn\\
&&\Omega_3=d\Omega_2+q^2\Omega_2\omega,\nn\\
&&....\nn\\
&&\Omega_k=d\Omega_{k-1}+q^{k-1}\Omega_{k-1}\omega.
\eea

\textbf{\emph{Remarks:}}
\begin{itemize}
\item
$\Omega_k=(\Omega_k)_i{}^j\in \Ld^k_q~~\forall i,j$ and $\Omega_k$ does not exist in dimension $n<k$.
\item

If for a particular $k$ we set $d^k=0, ~\Omega_k=0$ then a solution is given by a Maurer-Cartan form for $\M=GL(N)$ thus:
\bea
&&\Theta_{k-1}=d^{k-1}g~g^{-1},~~g\in GL(N),\nn\\
&&\Theta_1=\theta=dg~g^{-1},\nn\\
&&\Theta_{r}=\left\{
               \begin{array}{ll}
                 d\Theta_{r-1}+q^{r-1}\Theta_{r-1}~\theta,& 0\leq r< k \\
                 0, & r\geq k.
               \end{array}
             \right.
 \nn\\
\eea

\item On the other hand,  $d\Omega_k=0,~~q^k=1~~\Ra~~$ $\Omega_{k+r}=\Omega_k~\Omega_r,~~~~\Omega_{r_1k+r_2}=\Omega_k{}^{r_1}~\Omega_{r_2}$, ~where ~$r,r_1,r_2$ as well as $k$ are positive integers.

\item Finally for the case we are most concerned with, it can be checked that  $d^{k}=0,~~[k]_q=0$ implies Bianchi' 2nd identity
\bea
&&\label{bianchi}D\Omega_k=d\Omega_k-[\omega,\Omega_k]=d\Omega_k-\omega\Omega_k+\Omega_k\omega=0,\\
&&~~\Ra~~d~\txt{Tr}\Omega_k+(1-q^k)\txt{Tr}(\Omega_k\omega) = d~\txt{Tr}\Omega_k= 0,\nn\\
&&~~~~~~\txt{Tr}(\omega\Omega_k)=q^k\txt{Tr}(\Omega_k\omega).
\eea
Therefore, locally in $\M$, one may be able to write $\txt{Tr}\Omega_k=d^{k-1}U_1$; the ``constant'' numbers $z_k=\int\txt{Tr}\Omega_k$ may be used to characterize topological nontriviality on $\M$.

We also remark here that (\ref{bianchi}) has a solution of the form
\bea
&&\Omega_k=\mathcal{P} e^{\int\omega}~C_k ~\bar{\mathcal{P}} e^{-\int\omega},~~dC_k=0
\eea
and that a Lagrange multiplier type analytic continuation can be carried out as explained in \cite{akofor}. We will now consider the following alternative method of analytic continuation.
\end{itemize}

\section{The Source equation}
 Since $d^{k}=0,~~[k]_q=0$ implies
\bea
&&d\Omega_k-\omega\Omega_k+\Omega_k\omega=0,
\eea
we may drop this condition ($d^{k}=0,~[k]_q=0$) by replacing it with a physical current $J$ on the RHS thus

\bea
\label{gem}&&D\Omega_k=J_k,~~~~J|_{d^{k}=0,~[k]_q=0}=0,\nn\\
&&D\Omega^\ast_{n-k}=\td{J}_k,~~~~\td{J}|_{d^{k}=0,~[k]_q=0}=0,\\
&&\ast:\Ld^k_q\ra \Ld^{n-k}_q,~~(f+g)^\ast=f^\ast+g^\ast,~~~~\ast\ast:\Ld^k_q\ra \Ld^k_q,\nn
\eea
where $\ast$ is meant to be an analog of the Hodge dual.
When $\td{J}=J$ we may impose the duality condition $\Omega^\ast_{n-k}=\Omega_k$ which may be compared with an instanton solution.

One may define
$J:=(d\Omega'_k-\omega'\Omega'_k+\Omega'_k\omega')\al$, where $\al$ is an arbitrary complex function $\al:\M\ra \mathbb{C}$ and $\omega'$ is a known connection defined (supported and/or generated) by a field in some region of $\M$ ~~\cite{akofor}. That is, the domain of $\omega$ contains, and is bigger than that of $\omega'$ meanwhile $\omega'$ acts as the source of $\omega$. $\omega$ in turn may be seen to be simply an analytic continuation (or extension) of $\omega'$ via the equation $D\Omega_k=J_k$. This means that outside the support of $\omega'$ we must have $d\Omega_k-\omega\Omega_k+\Omega_k\omega=0$ ~~\cite{akofor}. In this dynamical picture, the condition $J|_{d^{k}=0,~[k]_q=0}=0$ is no longer necessary and may be completely eliminated. One may not be able to find solutions that satisfy these equations simultaneously for all $k$ and if that happens to be the case, then one may choose the equation(s) that best suit(s) a particular purpose.

\emph{\textbf{Remarks:}} The equations ($\ref{gem}$) are similar in form to those of electromagnetism and may therefore be interpreted analogously. The $q$ parameters and $d^k\neq 0$ conditions have each acquired physical significance through the introduction of the supposedly physical currents $J,\td{J}$.

\section{Illustrating q-symmetrization }\label{q-symm}
We wish to illustrate the index structure of $q$.
Consider the $q$-deformed 2-form $f$ of the gauge potential $A$ which may now be written as
\bea
&&f=dA+qA^2=(\del_\mu A_\nu+q_{\mu\nu}A_\mu A_\nu)dx^\mu dx^\nu\eqv f_{\mu\nu}dx^\mu dx^\nu,
\eea
where $dx^\mu$ are basis 1-forms such that $dx^\mu dx^\nu=q_{\nu\mu}dx^\nu dx^\mu,~~q_{\nu\mu}={1\over q_{\mu\nu}}$.
Then the appropriately symmetrized component form of $f$ is
\bea
&&F_{\mu\nu}={1\over 2}(f_{\mu\nu}+q_{\mu\nu}f_{\nu\mu})={1\over 2}\{\del_\mu A_\nu+q_{\mu\nu}A_\mu A_\nu+ q_{\mu\nu}(\del_\nu A_\mu+q_{\nu\mu}A_\nu A_\mu)\}\nn\\
&&~~~~={1\over 2}\{\del_\mu A_\nu+q_{\mu\nu}\del_\nu A_\mu+q_{\mu\nu}(A_\mu A_\nu+ q_{\nu\mu}A_\nu A_\mu)\}.\nn\\
&&F_{\mu\nu}=q_{\mu\nu}F_{\nu\mu}.
\eea

For electromagnetism (ie. an Abelian case), $A_\mu A_\nu=A_\nu A_\mu$~ and ~
\bea
F_{\mu\mu}={1+q_{\mu\mu}\over 2}(\del_\mu A_\mu+q_{\mu\mu}A_\mu A_\mu)=0~~\Ra~~q_{\mu\mu}=-1~~\forall \mu.
\eea
Therefore,
\bea
&&F_{\mu\nu}={1\over 2}\{\del_\mu A_\nu+q_{\mu\nu}\del_\nu A_\mu+q_{\mu\nu}(1+q_{\nu\mu})A_\mu A_\nu)\}.\nn\\
&&F^{\mu\nu}:={1\over 2}\{\del^\mu A^\nu+q_{\nu\mu}\del^\nu A^\mu+q_{\nu\mu}(1+q_{\mu\nu})A^\mu A^\nu)\}.\nn\\
&&F^{\mu\nu}=q_{\nu\mu}F^{\nu\mu}
\eea
and the Lagrangian for electrodynamics is
\bea
&&\mathcal{L}=F_{\mu\nu}F^{\mu\nu}-i\bar\psi \slash \!\!\!\! D\psi,~~\slash \!\!\!\! D=\gamma^\mu(\del_\mu+ieA_\mu).
\eea

In  $D=1+d,~~d\geq 2$ the Fourier space propagator
\bea
\Delta_{\mu\nu}(k)=(k^2\eta_{\mu\nu}+q_{\mu\nu}~k_\mu k_\nu)^{-1},
\eea
will be well defined if the $q$'s are chosen accordingly.

\subsection{The q-symmetrization in general}
Let \bea
&&dx^{i_1i_2...i_m}=dx^{i_1}dx^{i_2}...dx^{i_m},~~dx^{i_1i_2..i_{k-1}i_k..i_m}=q_{i_ki_{k-1}}dx^{i_1i_2..i_ki_{k-1}..i_m},\nn\\
&&q_{ji}={1\over q_{ij}},
\eea
 then
\bea
dx^{ijk}=q_{ki}q_{kj}dx^{kij}= q_{ki}q_{ji}dx^{jki}=q_{ji}dx^{jik}=q_{ji}q_{kj}q_{ki}dx^{kji}=q_{kj}dx^{ikj},
\eea
and therefore the symmetrized component form of
\bea
f=f_{ijk}dx^{ijk}
\eea
is
\bea
&&F_{ijk}=f^q_{ijk}={1\over 3!}\{f_{ijk}+q_{jk}q_{ik}f_{kij}+q_{ij}q_{ik}f_{jki}+q_{ij}f_{jik}+q_{ij}q_{ik}q_{jk}f_{kji}+q_{jk}f_{ikj}\}\nn\\
&&~~~~\eqv{1\over 3!}\sum_{p\in S_3}Q_{p;123}~f_{i_{p(1)}i_{p(2)}i_{p(3)}},\nn\\
&&f^q_{ijk}=q_{ij}f^q_{jik}
\eea
In general
\bea
&&f=f_{i_1i_2...i_m}dx^{i_1i_2...i_m},\nn\\
&&F_{i_1i_2...i_m}=f^q_{i_1i_2...i_m}={1\over m!}\sum_{p\in S_m}Q_{p;12...m}~f_{i_{p(1)}i_{p(2)}...i_{p(m)}},
\eea
where the rule for obtaining the $Q$'s is to first move the rightmost element, of the identity permutation, in the given permutation to its usual (rightmost) position in the identity permutation, while collecting any resulting $q$-factors with appropriate indices, and then the next rightmost, and next until the identity order of the permutation is restored. When trying to generate the $Q$'s corresponding to the permutation group $S_{m}$, it is useful to remember that elements of the $m$th order permutation group $S_m$ can be obtained from those of $S_{m-1}$, \emph{as a subset of $S_m$}, simply by a ``cycling'' operation $C_m$. That is,

\bea
&&S_{m}=(1\oplus C_m\oplus C_m{}^2\oplus...\oplus C_m{}^{m-1})S_{m-1}\nn\\
&&~~~~\eqv S_{m-1}\cup C_mS_{m-1}\cup C_m{}^2S_{m-1}\cup...\cup C_m{}^{m-1}S_{m-1},\nn\\
&&C_m: S_m\ra S_m,~~i_1i_2...i_m\mapsto  i_mi_1i_2...i_{m-1},~~~~C_m{}^m=1.
\eea
Thus obtaining the $Q$'s for $S_m$ is easier if the $Q$'s are known for $S_{m-1}\subset S_m$.

\section{Conclusion}
We have proposed a way to directly generate higher curvatures by replacing the $d^N=0,~q^N=1$ condition with a physical current. This has been motivated by the equivalence of the condition $d^N=0,~q^N=1$ to Bianchi's 2nd identity and also by the similarity between the proposed equations and Maxwell's equations for electromagnetism.

 \section{ACKNOWLEDGEMENTS}
This work was supported by the US Department of Energy under grant number DE-FG02-85ER40231.
\bibliographystyle{apsrmp}

 \ed